\documentclass[prl,twocolumn,showpacs,amsfonts,superscriptaddress,
amsmath,floatfix]{revtex4}
\usepackage{graphicx}
\usepackage{psfrag}
\newcommand{\Journal}[4]{#1 {\bf #2}, #3 (#4)}
\newcommand{\PR}{Phys. Rev.}
\newcommand{\PRL}{Phys. Rev. Lett.}
\newcommand{\PRA}{Phys. Rev. A}

\newcommand{\JMP}{J. Math. Phys.}

\newcommand{\Science}{Science}
\newcommand{\PLA}{Phys. Lett. A}
\begin{document}
\title {Bright and dark solitary waves in a one-dimensional spin-polarized gas of
fermionic atoms with p-wave interactions in a hard-wall trap}
\author{M. D. Girardeau}
\email{girardeau@optics.arizona.edu}
\affiliation{College of Optical Sciences, University of Arizona,
Tucson, AZ 85721, USA}
\author{E. M. Wright}
\email{ewan.wright@optics.arizona.edu}
\affiliation{College of Optical Sciences, University of Arizona,
Tucson, AZ 85721, USA}
\affiliation{Department of Physics, University of Arizona,
Tucson, AZ 85721, USA}
\date{\today}
\begin{abstract}
In this paper we elucidate the physics underlying the fact that both bright
and dark solitary waves can arise in a one-dimensional spin-polarized gas of
fermionic atoms with attractive three-dimensional p-wave interactions in a
hard-wall trap. This is possible since the one-dimensional fermion system
can be mapped to a system of bosons described by the Lieb-Linger model
with either repulsive or attractive
delta-function interactions which can support solitary waves.
\end{abstract}
\maketitle
Advances in experimental techniques for probing
ultracold gases have resulted in a shift in emphasis in theoretical
and experimental work in recent years from effective field approaches to more
refined methods capable of dealing with strong correlations. Such strong
correlations occur in ultracold gases
confined in de Broglie waveguides with transverse trapping
so tight that the atomic dynamics is essentially one-dimensional (1D)
\cite{Ols98}, with confinement-induced resonances \cite{Ols98,GraBlu04}
allowing Feshbach resonance tuning \cite{Rob01} of the effective 1D
interactions to very large values.  This has led to experimental verification
\cite{Par04Kin04,Kin05,Kin06} of the fermionization of bosonic ultracold
vapors in such geometries predicted by the Fermi-Bose (FB)
mapping method \cite{Gir60Gir65}, an exact mapping of a 1D gas of bosons with
point hard core repulsions, the ``Tonks-Girardeau'' (TG) gas, to an
\emph{ideal} spin-aligned Fermi gas. The ``fermionic Tonks-Girardeau'' (FTG)
gas \cite{GirOls03,GirNguOls04}, a 1D
spin-aligned Fermi gas with very strong \emph{attractive}
interactions, can be realized by a 3D p-wave Feshbach resonance as, e.g., in
ultracold $^{40}$K vapor \cite{Tik04}. It has been pointed out
\cite{GraBlu04,GirOls03,GirNguOls04} that the generalized FB mapping
\cite{CheShi98,GraBlu04,GirOls03,GirNguOls04} can be exploited to map the
\emph{ideal FTG} gas with \emph{infinitely strong} attractive
interactions to the \emph{ideal Bose} gas, leading to ``bosonization'' of many
properties of this Fermi system. In a recent paper \cite{GirWri07} we examined
how the properties of such an FTG gas on a mesoscopic ring are changed for an
even number $N$ of fermions when the strength of the atom-atom attraction is
made finite, in which case there exists an FB
mapping between the FTG gas and a system of $N$ bosons with repulsive
delta-function interactions.

It has been realized in recent
years that the FB mapping method used to exactly solve the TG gas
\cite{Gir60Gir65} and FTG gas \cite{GirOls03,GirNguOls04} is not restricted
to the TG case of point hard core boson-boson repulsion and the ideal FTG case
of infinite zero-range fermion-fermion attraction. In fact, the same unit
antisymmetric mapping function used there,
$A(x_1,\cdots,x_N)=\prod_{1\le j<\ell\le N}\text{sgn}(x_j-x_\ell)$,
provides an exact mapping between bosons with delta-function repulsions
$g_{1D}^B\delta(x_j-x_\ell)$ of any strength, i.e., the Lieb-Liniger (LL)
model \cite{LL63}, and spin-aligned fermions with attractive interactions
of a generalized FTG form with reciprocal fermionic coupling constant
$g_{1D}^F$; here $\text{sgn}(x)$ is $+1\ (-1)$ if $x>0\ (x<0)$. All previous
work, including \cite{GirWri07} and a recent Bethe ansatz treatment of such
a system in a hard-wall trap \cite{HZC07}, have considered only the case
where the strongly attractive spin-aligned Fermi gas maps to the LL model
with \emph{repulsive} interactions, corresponding to a \emph{negative} 1D
scattering length $a_{1D}$. This restriction is not necessary
and a minor change in the parameters results in a mapping of the
FTG gas to an LL model with
\emph{attractive} interactions, corresponding to a \emph{positive} scattering
length $0<a_{1D}<\infty$, an equally physical regime. The \emph{ideal} FTG
gas $|a_{1D}|\to\infty$ lies on the ``knife edge'' between the attractive
and repulsive LL regimes, where $a_{1D}$ jumps discontinuously from $-\infty$
to $+\infty$, as might have been expected from the fact that it corresponds
to a zero-energy scattering resonance. The purpose of the present paper
is to elucidate how this works, and to compare and contrast
the ground states of the spin-aligned Fermi gas in these two regimes. In particular,
for the case of a hard-wall trap and using the mapping between the FTG and LL
model, we expose that both bright and dark solitary waves can arise.
The possibility of solitons
in degenerate Fermi gases \cite{GirWri00,Karpiuk02,WitBre05} has certainly been investigated
previously, and the novelty in the present work is that it includes the p-wave
fermion-fermion  interactions, highlights the relation between the Fermi and Bose systems,
and exposes the role played by the solitary waves of the underlying Bose system.

{\it 3D p-wave resonance and induced 1D interaction:} 3D p-wave Feshbach
resonances have been observed in some species of ultracold gases, e.g.,
in $^{40}$K \cite{Tik04}. When such a gas is contained in a de Broglie
waveguide with tight transverse trapping so as to reach the regime of
effectively 1D dynamics, such a 3D resonance leads to a nearby
confinement-induced resonance (CIR) in the 1D scattering, and it has been
shown \cite{GraBlu04,GirOls03,GirNguOls04} that the 3D p-wave scattering
length $a_p$ and 1D scattering length $a_{1D}$ in the neighborhood of such a
resonance have the connection
\begin{eqnarray}\label{renorm}
a_{1D}&=&\frac{6V_{p}}{a_{\perp}^2}[1+12(V_{p}/a_{\perp}^3)
|\zeta(-1/2,1)|]^{-1} ,
\end{eqnarray}
where $a_{\perp}=\sqrt{\hbar/m_\text{red}\omega_{\perp}}$ is the transverse
oscillator length, $V_{p}=a_{p}^{3}=-\lim_{k\to
0}\tan\delta_{p}(k)/k^3$ is the p-wave ``scattering volume'',
$a_{p}$ is the p-wave scattering length,
$\zeta(-1/2,1)=-\zeta(3/2)/4\pi=-0.2079\ldots$ is the Hurwitz zeta
function evaluated at $(-1/2,1)$, and $m_\text{red}=m/2$ is
the reduced mass. The expression (\ref{renorm}) has a resonance at
a \emph{negative} critical value
$V_{p}^{crit}/a_{\perp}^{3}=-0.4009\cdots$, implying that the CIR only occurs
when $a_p<0$. However, by varying an external magnetic field so as to sweep
$a_p$ through the critical value, one can in principle vary $a_{1D}$ from
$-\infty$ to $+\infty$. The ideal FTG gas, which maps to the ideal Bose gas,
correspond to the point where $a_{1D}$ jumps discontinuously from $-\infty$
to $+\infty$, and nearby values of $a_p$ correspond to an
``imperfect FTG gas'', which we shall call herein simply an FTG gas.
$a_{1D}$ is defined implicitly by the contact condition
\cite{GraBlu04,GirOls03,GirNguOls04}
\begin{equation}\label{Fermi-contact}
\Psi_{F}(x_{12}=0+)=-\Psi_{F}(x_{12}=0-)
= -a_{1D}\Psi_{F}^{'}(x_{12}=0\pm)
\end{equation}
where $\Psi_{F}(x_{12})$ is the fermionic relative wave function and
the prime denotes differentiation.

\begin{figure}[t]
  \centering
\includegraphics[width=7.5cm,angle=0]{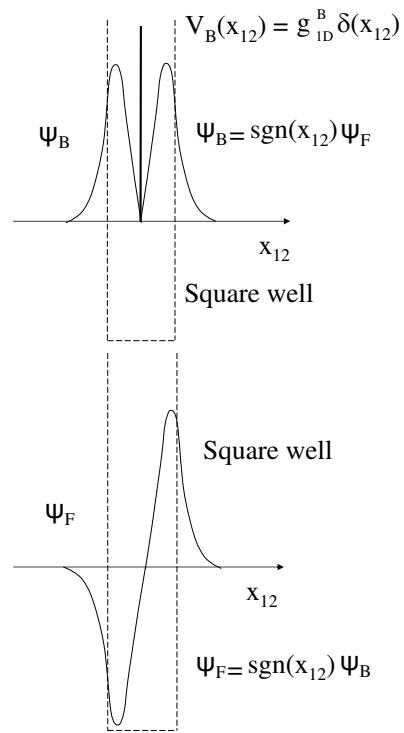}
  \caption{Two-particle fermionic relative wave function $\Psi_F$ and mapped
bosonic wave function $\Psi_B$ as a function of $x_{12}=x_1-x_2$,
for the case where the negative potential energy slightly exceeds the positive
kinetic energy, so the strongly attractive Fermi system maps
to a Lieb-Liniger Bose gas with \emph{weak attractions}.
The potential for both $\Psi_F$ and $\Psi_B$ is a deep and narrow square
well plus a point hard core, but in the zero-range limit its effect on $\Psi_B$
outside the well is the same as that of $V_B=g_{1D}^B\delta(x_{12})$,
where here $g_{1D}^B<0$.}
  \label{fig1}
\end{figure}

Several different 1D pseudopotentials have been used in the literature to
represent a 1D potential leading to this contact condition, but most of these
involve combinations of delta functions and derivatives
which make them physically opaque. However, the physics is clearly
exhibited if one instead uses a suitable zero-range limit of a deep and
narrow square well of depth $V_0$ and width $2x_0$, with the zero-range
limit $V_0\to\infty$ and $x_0\to 0$ carried out at constant $V_0x_0^2$
\cite{GirOls03,GirNguOls04,GirWri07}. Consider first the two-body problem
$N=2$ in infinite space. The relative wave function $\Psi_{F}(x_{12})$
of the ground state inside
the well is $\sin(\kappa x_{12})$ where the value of the positive parameter
$\kappa$ is determined by the total energy (potential plus kinetic), or
equivalently, by $a_{1D}$. One finds
$\kappa x_0=\frac{\pi}{2}+\frac{2x_0}{\pi a_{1D}}$ as $x_0\to 0$
\cite{GirOls03,GirNguOls04}. The ideal FTG gas (total energy zero, i.e.,
zero-energy resonance) corresponds to infinite $a_{1D}$ and hence
$\kappa x_0=\frac{\pi}{2}$. The imperfect FTG gas studied in \cite{GirWri07}
has negative $a_{1D}$ and positive total energy, i.e., inside the well
the positive kinetic energy exceeds the negative potential energy $-V_0$
by a finite amount. What we wish to point out here is that exactly the
same relation $\kappa x_0=\frac{\pi}{2}+\frac{2x_0}{\pi a_{1D}}$ also
applies to the case where $a_{1D}$ is positive and the negative potential
energy exceeds the positive kinetic energy inside the well, resulting
in a finite and negative total energy, i.e., a bound ground state.
The generalized FB mapping \cite{CheShi98,GraBlu04,GirOls03,GirNguOls04}
$\Psi_{B}(x_{12})=\text{sgn}(x_{12})\Psi_{F}(x_{12})$ maps the fermionic
wave function to a bosonic one $\Psi_{B}$ with the same scattering length
$a_{1D}$. The case $a_{1D}<0$ of this mapping was illustrated in Fig. 1
of \cite{GirWri07}, and the case $a_{1D}>0$ which we wish to study here
is illustrated by the present Fig.\ref{fig1}. For consistency of the mapping
one should add a zero-diameter hard core at $x_{12}=0$ to the potential,
which has no effect on $\Psi_{F}$ \cite{GirWri07}.
In the zero-range limit the effect of this potential on $\Psi_{B}$
outside the well is exactly the same as that of an LL delta function
interaction $V_B=g_{1D}^B\delta(x_{12})$ with
$g_{1D}^B=-2\hbar^2/ma_{1D}<0$. The corresponding fermionic coupling
constant is reciprocally related to $g_{1D}^B$, i.e.,
$g_{1D}^F=-2\hbar^2a_{1D}/m<0$ and $g_{1D}^Bg_{1D}^F=\frac{4\hbar^4}{m^2}$
\cite{Note1}.
Outside the well the fermionic and mapped bosonic wave functions are both
decaying exponentials, i.e., the ground state is bound.
In the case of enclosure in a longitudinal
box or well, or trapping on a ring, the wave functions are modified at
large distances in accordance with the boundary conditions and/or trap
potential, but the above represents the exact behavior of the wave functions
as $x_{12}\to 0$.

{\it Many-fermion ground state:} For a system of $N$ fermions the Fermi wave
function may be related to the underlying Bose wave function via the mapping
$\Psi_F=A\Psi_B$, where $\Psi_B$ is the ground state of the N-boson LL Hamiltonian.
The relative strengths of the induced 1D fermion interactions and the LL model
interactions are quantified in the dimensionless coupling
coefficients $\gamma_F=\frac{mg_{1D}^Fn}{\hbar^2}$ and
$\gamma_B=\frac{mg_{1D}^B}{\hbar^2 n}$, respectively, where $n=\frac{N}{L}$ is
the linear atomic density, and $\gamma_F\gamma_B=4$.  The ideal FTG gas is
realized in the limit $a_{1D}\to-\infty$ so that $\gamma_F\to\infty$ and
$\gamma_B\to 0$.  Here we are interested in the limit of finite but very
large $|a_{1D}|$, implying $|\gamma_F|>>1$ and
$|\gamma_B|<<1$.  This is precisely the limit where the ground state
properties of the LL Hamiltonian can be accurately captured using mean field
theory in which all N bosons are assumed to occupy the same normalized single particle
orbital $\phi$ that is determined as the ground state solution of the
Gross-Pitaevskii equation (GPE) \cite{LifPit89}
\begin{equation}
\mu\phi = -\frac{\hbar^2}{2m}\frac{d^2\phi}{dx^2} + g_{1D}^B(N-1)|\phi|^2\phi  ,
\label{GPE}
\end{equation}
with $\mu$ the chemical potential, and the GPE is to be solved subject to the hard-wall
boundary conditions $\phi(0)=\phi(L)=0$ for a trap of length L. (We employ a hard-wall trap
due to the fact that the solutions of Eq. (\ref{GPE}) are well known, but our
general findings would
also apply to a harmonic trap).  The $N$-particle fermion
wave function can then be written as $\Psi_{F}(x_{1},\cdots,x_{N})=
A(x_{1},\cdots,x_{N})\prod_{j=1}^{N}\phi(x_{j})$, and the corresponding reduced
one-body density matrix
(OBDM) may be expressed as $\rho_{1}(x,x')=N\phi(x)\phi^*(x')[F(x,x')]^{N-1}$ where
$F(x,x')=\int_{-\infty}^{\infty}\text{sgn}(x-x'')\text{sgn}(x'-x'')
|\phi(x'')|^2 dx''$ \cite{BenErkGra05}.  The single-particle momentum spectrum
$n(k)$
for the Fermi gas may be obtained by Fourier transformation of the OBDM \cite{HZC07},
and whereas the density
$\rho(x)=\rho_{1}(x,x)=N|\phi(x)|^2$ is the same for both the Fermi and mapped Bose systems,
the momentum spectra of the two systems differ greatly \cite{HZC07}.

{\it $a_{1D}<0$:} In this case
$g_{1D}^F>0$ and the mapped LL boson interactions
are repulsive, $g_{1D}^B>0$ \cite{Note1}.  The ground state solution of
the GPE
(\ref{GPE}) with hard-wall boundary conditions is a dark solitary wave whose
spatial form is an
elliptic Jacobi snoidal function: Exhaustive details of the form of the
solutions are given in Ref. \cite{CarClaRei00a}.  In the limit of a single atom
the GPE ground state degenerates into a half period of a sine function
over the length $L$ of the trap.  In contrast, for
a large number of atoms $N>>1$ the solution tends to the value $\sqrt{n}$ within the center
of the trap, but approaches the classic hyperbolic tangent form of a dark soliton
near the hard wall boundaries within a distance characterized by the healing
length $\xi_h=\sqrt{|a_{1D}|/(2n)}<<L$ which follows from the relation
$\hbar^2/(2m\xi_h^2)=g_{1D}^B n$ \cite{CarClaRei00a}.

\begin{figure}[t]
  \centering
\includegraphics[width=7.5cm,angle=0]{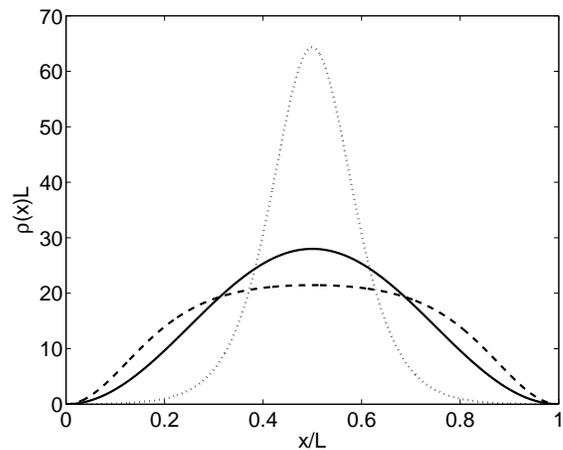}
  \caption{Scaled density profile $\rho(x)L$ versus scaled coordinate $x/L$ for
  the ideal FTG (solid line),
 finite $a_{1D}<0$ (dashed line), and finite $a_{1D}>0$ (dotted line).
 The results are for $N=14$ and $|\gamma_B|=0.1$.}
  \label{fig2}
\end{figure}

Figure \ref{fig2} shows the calculated scaled density profile $\rho(x)L$ for $N=14$ fermions and
$\gamma_B=0.1$ for the case $a_{1D}<0$ (dashed line) and also
the ideal FTG gas (solid line) and we see that the density profile is broadened by the
finite fermion interactions.  For the ideal FTG gas the competing effects of the
infinitely strong attractive 1D interactions and repulsion due to Fermi degeneracy pressure
exactly cancel and that is why the ideal FTG maps to the ideal Bose gas with no residual
interactions.
In contrast, for the case considered here with finite but large $a_{1D}<0$,
the Fermi degeneracy pressure dominates over the attractive interactions and
that is why the FTG is mapped to a LL model with repulsive interactions that broaden the
density profile.  This broadening of the density profile in the presence of decreasing
attractive induced 1D fermion interactions was first found by Hao, Zhang, and
Chen \cite{HZC07} using exact solutions for the fermion system based on the Bethe ansatz.
Here we have elucidated the underlying physics by exposing the relation to the mapped LL
model with repulsive interactions and its dark solitary wave solution.

\begin{figure}[t]
  \centering
\includegraphics[width=7.5cm,angle=0]{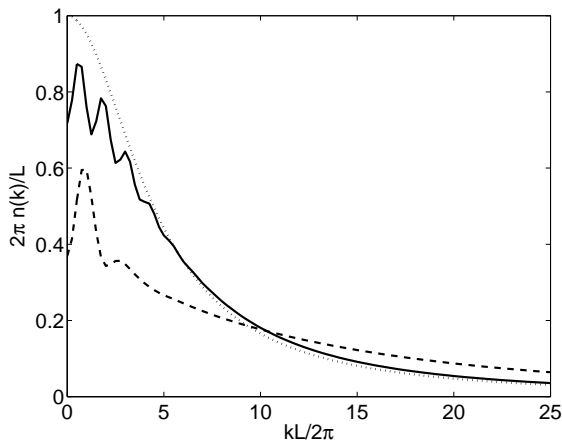}
  \caption{
Scaled momentum spectrum $2\pi n(k)/L$ versus scaled
momentum $kL/2\pi$ for $|\gamma_B|=0.1$, $N=14$ fermions, $a_{1D}<0$
(solid line), $a_{1D}>0$ (dashed line),
and using Eq. (\ref{nk}) (dotted line).
}
  \label{fig3}
\end{figure}

Figure \ref{fig3} shows the scaled momentum spectrum $2\pi n(k)/L$ versus
scaled momentum $kL/2\pi$ for $|\gamma_B|=0.1$, $N=14$ fermions,
and $a_{1D}<0$ (solid line).  Bender, Erker, and Granger \cite{BenErkGra05}
have previously shown that for the ideal FTG the high momentum
tails of the momentum spectrum are intimately related to the decay of the OBDM $\rho_1(x,x')$
as one moves off-diagonal $x\ne x'$.  In the thermodynamic limit they found that the decay of
the correlations is exponential with decay constant $2n$, and in our
notation the scaled momentum spectrum becomes
\begin{equation}\label{nk}
\frac{2\pi n(k)}{L} = \frac{4n^2}{(4n^2+k^2)}  .
\end{equation}
Thus, the larger the density $n$ is the more rapidly the correlations decay,
and the larger the width $\Delta k=4n$ of the momentum spectrum is.
The dotted line in Fig. \ref{fig3} shows the momentum spectrum using
Eq. (\ref{nk})
and yields reasonable agreement with the high momentum tails in the
numerical solution (solid line) even though
we have a finite trap and are not in the thermodynamic limit.

{\it $a_{1D}>0$:} In this case $g_{1D}^F<0$ and the mapped LL boson
interactions
are attractive, $g_{1D}^B<0$ \cite{Note1}.  The ground state solution of
the GPE (\ref{GPE})
with hard-wall boundary conditions is a bright solitary wave whose spatial
form is an
elliptic Jacobi cnoidal function: Details of the form of the
solutions are given in Ref. \cite{CarClaRei00b}.  In the limit of a single atom
the ground state solution degenerates into a half period of a sine function
over the length $L$ of the trap, whereas for
a large number of atoms $N>>1$ the solution becomes a localized peak at the center
of the trap and approaches the classic hyperbolic secant form of a bright soliton
with a characteristic width $w<L$ determined by
$\hbar^2/(2mw^2)=|g_{1D}^B|N/w$ \cite{CarClaRei00b}.  Thus, for a large number of atoms
the bright soliton width becomes independent of the trap length $L$,
$w=a_{1D}/(4N)$.

Figure \ref{fig2} shows the calculated scaled density profile $\rho(x)L$ for
$N=14$ fermions and $|\gamma_B|=0.1$ for finite $a_{1D}>0$ (dotted line) and
also
the ideal FTG gas (solid line) and we see that the density profile is narrowed.
This can be intuited using the fact that the mapped LL model
has attractive interactions so that the density profile is narrowed compared to
the free Bose gas, which maps to the ideal FTG.

Figure \ref{fig3} shows the
scaled momentum spectrum $2\pi n(k)/L$ versus
momentum $kL/2\pi$ for $|\gamma_B|=0.1$, $N=14$
fermions, and $a_{1D}>0$ (dashed line), and by comparison with the case
$a_{1D}<0$ (solid line) the momentum spectrum is broader.
This may be understood as follows:
Inspection of
the plots (not shown) of the OBDM $\rho_1(x,x')$ for $a_{1D}>0$
shows that it decays significantly more
rapidly as one moves off-diagonal $x\ne x'$ in comparison to the case
$a_{1D}<0$, which implies that the momentum spectrum
will be broader \cite{BenErkGra05}, see the discussion surrounding Eq. (\ref{nk}).
This is also consistent with the fact that the peak density
of the bright solitary wave is higher than
that for the dark solitary wave, see Fig. \ref{fig2}, and the
momentum spectrum width $\Delta k$ increases with density.

In summary, we have shown that both bright
and dark solitary waves can arise in a one-dimensional spin-polarized gas of
fermionic atoms with attractive 3D p-wave interactions in a hard-wall trap, by 
virtue of the fact that the fermion system can be mapped to a system of 
interacting bosons with induced 1D interactions that can be either attractive 
or repulsive, exposing the underlying solitary waves.

%
%

%
\end{document}